\documentclass[prb,aps,twocolumn,floatfix,amsmath,amssymb,superscriptaddress,tightenlines]{revtex4}
\usepackage{graphicx}% Include figure files
\usepackage{epstopdf}
\newcommand{\be}{\begin{equation}}
\newcommand{\ee}{\end{equation}}
\usepackage{amsmath}
\usepackage{amsfonts}
\usepackage{bm}
\usepackage{color}

\usepackage{amsthm}

\newtheorem{definition}{Definition}

\begin{document}
\title{Topological Order at Non-zero Temperature}
\author{Matthew B. Hastings}
\affiliation{Duke University, Department of Physics, Durham, NC, 27708}
\affiliation{Microsoft Research, Station Q, CNSI Building, University of California, Santa Barbara, CA, 93106}

\begin{abstract} 
We propose a definition for topological order at nonzero temperature in analogy
to the usual zero temperature definition that
a state is topologically ordered, or ``nontrivial", if it cannot be transformed into a product state (or a state close to a product state) using a local (or approximately local) quantum circuit.  We prove that any two dimensional Hamiltonian which is a sum of commuting local terms is not topologically ordered at $T>0$.  We show that
such trivial states cannot be used to store quantum information using certain stringlike operators.  This definition is not too restrictive, however, as the four dimensional toric code does have a nontrivial phase at nonzero temperature.
\end{abstract}
\maketitle
Topological quantum computation\cite{tqc1,tqc2} is one of the most promising approaches for building a quantum computer.  The nonlocal encoding of quantum information in the ground state subspace of topologically ordered systems protects it against decoherence.  Topological order is also one of the most interesting current topics in condensed matter physics, as models such as the Levin-Wen models\cite{lw} provide a method for constructing novel phases.

One way of characterizing the nontrivial properties of these phases is the following definition: a state is ``trivial" if it is a product state, or if it can be transformed into a state which approximates a product state using a local or approximately local quantum circuit.  A state is nontrivial, or topologically ordered, otherwise.  Despite the usefulness of this definition in terms of circuits, we do not have a similarly useful definition for states at temperature $T>0$.  

Here, we propose an analogous ``circuit definition" for topological order at $T>0$.  
Another method used previously to study topological order at $T>0$ is topological mutual information\cite{topomi,topomi2}, a generalization of topological entanglement entropy\cite{topoee1,topoee3}.  
While a nonzero topological mutual information for $T$ near zero is a useful numerical signature of nontrivial phases\cite{isakov}, the use of topological entropy can give different answers from the circuit definition even for at $T=0$\cite{diff}, so both definitions are useful to fully characterize a state.
Also, it has been shown that in many two-dimensional theories there exist
stringlike logical operators\cite{defect1,defect2}.  If these logical operators can be generated by dragging defects with energy $O(1)$, then, 
since there is a nonzero density of such defects at $T>0$, the information which would be topologically protected at zero temperature will decohere rapidly at $T>0$ (though we emphasize that the defect density is exponentially small in $1/T$; as an experimental example, fractional Hall conductance is accurately quantized\cite{hallcond} even at $T>0$).  While this kind of ``operational definition" is practically useful, the circuit definition here  will lead to similar operational results for {\it all} Hamiltonians which are a sum of local commuting terms.

Conversely, we will show that the four dimensional toric code\cite{4dtoric} has topological order under the circuit definition for sufficiently small $T>0$ (this will not be a rigourous proof since we will heuristically argue for the existence of certain operators using previous results, but we then prove that the existence of these operators is inconsistent with a trivial state).  One advantage of a definition using the density matrix rather than the excitation above the ground state is seen in a toy system in the Appendix which is topologically trivial at $T=0$ but non-trivial at some small $T>0$.

{\it Topological Order at Zero Temperature---}
We start with various previous definitions of topological order at $T=0$. 
Topologically ordered Hamiltonians according to one definition have a ground state degeneracy that depends upon the topology of the manifold, and the ground state subspace obeys a property called the ``disk axiom" or ``TQO-1" \cite{tqc1,tqc2,disk1,bhm,bh}, which
we quantify by:
\begin{definition}
Let $L$ denote the system size and let $P$ denote the projector onto the ground state subspace.  
Let $L^*$ denote some length smaller than $L$.  
Then, the ground state subspace has $(L^*,\epsilon)$ topological degeneracy\cite{bhv} if for any operator $O$ supported on a set of diameter smaller than $L^*$ there is a scalar $z$ such that
\be
\Vert POP-z P\Vert \leq \epsilon.
\ee
\end{definition}

Using circuits we can also define topological order for systems with a unique ground state.
Consider a unitary quantum circuit $U$ where the depth of the circuit multiplied by the maximum range of each unitary in the circuit is bounded by some range $R$.  
Note that for any operator $O$ supported on a set $Z$, $U^\dagger O U$ is supported on the set of sites within distance $R$ of $Z$.
\begin{definition}
Let $\psi_0$ be the ground state of the Hamiltonian $H$.
We say that the state $\psi_0$ is $(R,\epsilon)$ trivial if there exists
a
unitary quantum circuit $U$ with range $R$ such that $|\psi_0- U \psi_{prod}|\leq \epsilon$ for some product state $\psi_{prod}$.
\end{definition}

Certainly, every state is $(L,0)$ trivial, so we are only interested in the case of $R<L$.
Colloquiually speaking a system will be trivial if it is $(R,\epsilon)$ trivial for some $\epsilon<<1$ and for some $R<<L$.
We can relate the two definitions\cite{bhv}: if a system has $(L^*,\epsilon)$ topological degeneracy, then no state $\psi_0$ in the ground state subspace is $(R,\delta)$ trivial for $R<L^*/2$ and sufficiently small $\delta,\epsilon$.
To see this, suppose $\psi_0$ is $(R,\delta)$ trivial.  Then, the expectation of any operator $O$ in state $\psi_{prod}$ is close to the expectation of $\langle \psi_0|U O U^\dagger|\psi_0\rangle$.  If $O$ is supported on a single site, $UOU^\dagger$ is supported on a set of diameter less than $L^*$ and so for such $O$, $\langle \psi_{prod}|O|\psi_{prod}\rangle \approx \langle U^\dagger \psi_1 | O | \psi^\dagger \psi_1 \rangle$.  That is, the subspace spanned by $\psi_{prod}$ and $U^\dagger \psi_1$ has $(1,\eta)$ topological degeneracy for $\eta={\cal O}(\epsilon+\delta)$.  However, for $\eta$ sufficiently small compared to inverse system size, no such state $U^\dagger \psi_1$ exists since $\psi_{prod}$ is a product state.

{\it Topological Order at Nonzero Temperature---}
We begin by defining a ``classical state of range $R$" $\rho_{cl}$ (this will replace the use of a product state when we define $T>0$ topological order)
to be a state such that
$\rho_{cl}=Z^{-1} \exp(-H_{cl})$
where $Z$ is a normalization factor and $H_{cl}$ is a Hamiltonian which is a sum of terms all acting on sets of diameter
at most $R$ and all of which are
diagonal in a product basis.

\begin{definition}
A density matrix $\rho$ is $(R,\epsilon)$ trivial if it is 
possible to
tensor in additional degrees of freedom ${\cal K}_i$ on each site, defining an enlarged space with Hilbert space ${\cal H}_i \otimes {\cal K}_i$ on each site, such that
\be
\label{trivdens}
|\rho-{\rm Tr}_{\{{\cal K}_i\}}\Bigl( U \rho_{cl} U^\dagger \Bigr)| \leq \epsilon,
\ee
where the $|...|$ denotes the trace norm (the trace norm of a Hermitian operator is the sum of the absolute values of its eigenvalues),
where $U$ is a
unitary quantum circuit
with range $R$, and $\rho_{cl}$ is a classical state of range $R$ (both $U$ and $\rho_{cl}$ are defined on the enlarged space), and
where the trace is over the added degrees of freedom ${\cal K}_i$.
\end{definition}
If we allow $H_{cl}$ to be unbounded and we allow the dimension of ${\cal K}_i$ to be unbounded, then this definition is equivalent to saying that $\rho$ is $(R,\epsilon)$ trivial if it is, up to error $\epsilon$ in trace norm, equal to an incoherent sum of $(R,0)$ trivial states: $\rho=\sum_a P(a) |\psi_{triv}(a) \rangle\langle \psi_{triv}(a)|$, for some probability distribution $P(a)$.
However, for some purposes one might want to  construct $\rho$ using a bounded $H_{cl}$ and a bounded dimension on ${\cal K}_i$.

{\it Absence of Topological Order For Two Dimensional Hamiltonians With Commuting Terms---}
We now show absence of topological order for any $T>0$, under the above definition, for two dimensional Hamiltonians which are a sum of commuting projectors.  The proof is based on showing that the density matrix can be approximately written as a weighted sum over density
matrices of a system with ``holes" in it as explained below, and then using results from \cite{bv} to write each such density matrix as a trivial state.

Consider a two dimensional Hamiltonian $H=\sum_X Q_X$, where the terms $Q_X$ are commuting projectors.  Assume that the terms in the Hamiltonian are local, so that each projector $Q_X$ is supported on some set $X$ which has diameter $R_{int}$ which is $O(1)$.  Further suppose that each site is in at most $O(1)$ of such sets $X$, and for simplicity consider a square lattice. 

The density matrix is $\rho=Z^{-1} \exp(-\beta H)$, where $Z$ is a normalization and $\beta=T^{-1}$.  Note that for any projector $Q_X$ we have
$\exp(-\beta Q_X)
= \sum_{s_X\in \{0,1\}} 
\Bigl( (1-s_X) \exp(-\beta) \cdot I + s_X (1-\exp(-\beta)) \cdot (I-Q_X) \Bigr)$
where we introduce an additional variable $s_X$, and sum over $s_X=0,1$ (this variable $s_X$ is unrelated to any local spin degrees of freedom of the Hamiltonian).
Thus, 
\be
\label{splitdefn}
\rho=\sum_{\{s_X\}} \frac{Z(\{s_X\})}{Z} \rho(\{s_X\}) P(\{s_X\}),
\ee
where the sum is over a set of variables $s_X$, each variable taking values $0$ or $1$,
with $P(\{s_X\})=\prod_X \bigl((1-s_X)\exp(-\beta)+s_X (1-\exp(-\beta)\bigr)$,
 and
where
\be
\label{sumup}
\rho(\{s_X\})\equiv
Z(\{s_X\})^{-1} \prod_X
\Bigl( (1-s_X) I + s_X (I-Q_X) \Bigr),
\ee
with
$Z(\{s_X\})^{-1}$ being a normalization.  Note that $\rho(\{s_X\})$ is maximally mixed on the ground state subspace of $H(\{s_X\})\equiv \sum_X s_X Q_X$.

We will show that the density matrix $\rho$, for any given $\beta$, is dominated by a sum over choices of $s_X$ in which there are lots of ``holes" in the lattice, where a hole corresponds to a disk $Y$ with radius greater than $R_{int}$,  such that for any $X$ with $X\cap Y \neq \emptyset$, we have $s_X=0$.  See Fig.~1.  (Note that the variables $s_X$ for $X$ not intersecting such a disk $Y$ are not determined by the choice of holes, these variables may be either $0$ or $1$).  We divide the square lattice into large squares of linear  size $l_{\beta}$, with $l_{\beta}$ exponentially large in $\beta$ as given in Eq.~(\ref{lbeta}) and logarithmically large in system size.  We call a configuration ``valid" if there is at least one hole per square.
We will show that for valid configurations the density matrix $\rho(\{s_X\})$
can be expressed as a local unitary with range $R=2 l_\beta$ acting on a classical state
and we bound the contribution to $\rho$ from invalid configurations.
Combining these results implies
that the sum over $s_X$ can be approximated by a local unitary acting on a classical state (the error $\epsilon$ arises from
the invalid configurations of $s_X$).

\begin{figure}
\label{fighole}
\includegraphics[width=1.9in]{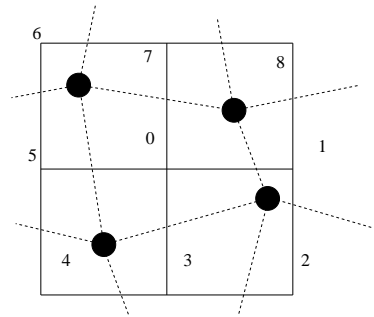}
\caption{Illustration of holes in the lattice.  The solid lines indicate the division of the lattice into squares of size $l_\beta$, and only some of the squares are illustrated.  Solid circles indicate holes in the lattice, with one hole per square.  Dashed lines connect holes (dashed line extending outside the solid lines connecting to only one hole are intended to indicate connections to holes in other square which are not illustrated).  Symbols $0,1,2,...$ indicate different subsets of the lattice with holes in it, with the dashed lines indicating divisions between those subsets.
}
\end{figure}

Consider a given valid choice of $s_X$.  
For such a valid choice of $s_X$, we draw an image such as in
Fig.~1.  As mentioned, the terms $s_X$ may be zero even if $X$ does not intersect a hole.  If a given choice of $s_X$ has more than one hole per square, we only indicate one hole in the square, choosing which one to take according to any arbitrary rule (for example, the hole closest to the top left corner).  We draw dashed lines connecting the holes.  The dashed lines break the lattice outside the holes into regions labelled $0,1,2,...$ as shown.
Then, 
$H(\{s_X\})$ can be re-written as
\be
H(\{s_X\})=\sum_{<a,b>} H_{a,b}+\sum_a H_a,
\ee
where $a$ and $b$ label regions bounded by dashed lines, the sum is over neighboring $a,b$, and
$H_{a,b}$ is supported on regions $a$ and $b$ (define $H_{a,b}=H_{b,a}$) and $H_a$ is supported on region $a$.  

So, by coarse-graining the lattice, the Hamiltonian is a sum of commuting terms, each acting on at most two regions.
For any $a$, the terms $H_{a,b}$ and $H_a$ which are supported on that region all commute with each other.  This allows us to
find a decomposition of the Hilbert space which greatly simplifies the description of the problem.
Eqs.~(\ref{decomp1},\ref{decomp2}) below differ slightly from
lemma 8 of \cite{bv}, but can be proven using the same tools of interaction algebra\cite{intalg} (see also \cite{qbp}):
decompose $H_{a,b}$  as a sum of product operators $H_{a,b}=\sum_\gamma O_a^{ab}(\gamma) O_b^{ab}(\gamma)$, where the operators $O_a^{ab}(\gamma),O_b^{ab}(\gamma)$ 
are supported on $a,b$ respectively and the operators $O_b^{ab}(\gamma)$ are chosen from an orthonormal basis.
 Then, $[O_a^{ab}(\delta),O_a^{ac}(\gamma)]=0$ for 
$b\neq c$, for all $\delta,\gamma$.  
Let ${\cal A}^{ab}$ be the algebra generated by the set of $O_a^{ab}(\gamma)$ for given $b$.  The algebras ${\cal A}^{ab}, {\cal A}^{ac}$ commute for $b \neq c$.  

Let ${\cal H}_a$ denote the Hilbert space on region $a$.  
One way for the two algebras to commute is simply that ${\cal H}_a$ decomposes into a tensor product of Hilbert spaces, and each ${\cal A}^{ab}$ acts on a different space.  However, this is not the only possibility.  Suppose, for example, the Hamiltonians $H_{a,b}$ are all diagonal in some product basis.  For example, consider an Ising Hamiltonian with all terms involving only operators $S^z$.  Then all the $H_{a,b}$ would commute, but we would not have this tensor product decomposition.  However,
we can decompose ${\cal H}_a$ into a direct sum of Hilbert spaces ${\cal H}_a^{\alpha(a)}$, and then further decompose each such Hilbert space ${\cal H}_a^{\alpha(a)}$ into a tensor product of spaces ${\cal H}_{a\rightarrow b}^{\alpha(a)}$ giving
\begin{eqnarray}
\label{decomp1}
{\cal H}_a& = &\bigoplus_{\alpha(a)} {\cal H}_a^{\alpha(a)}
= \bigoplus_{\alpha(a)} \Bigl( {\cal H}_{a,a}^{\alpha(a)} \otimes \bigotimes_{<b,a>} {\cal H}_{a\rightarrow b}^{\alpha(a)} \Bigr),
\end{eqnarray}
where the product is over $b$ that neighbor $a$,
such that each operator $H_{a,b}$ can be decomposed as
\be
\label{decomp2}
H_{a,b}=\sum_{\alpha(a),\beta(b)} P_a^{\alpha(a)} P_b^{\beta(b)} H_{a,b}^{\alpha(a),\beta(b)},
\ee
where $P_a^{\alpha(a)}$ is the operator on ${\cal H}_a$ which projects onto ${\cal H}_a^{\alpha(a)}$ and $H_{a,b}^{\alpha(a),\beta(b)}$ acts on
the subspace of ${\cal H}_a^{\alpha(a)} \otimes {\cal H}_b^{\beta(b)}$ given by ${\cal H}_{a\rightarrow b}^{\alpha(a)} \otimes {\cal H}_{b \rightarrow a}^{\beta(b)}$.

Define $Q_{a,b}$ to project onto the nonzero energy states of $H_{a,b}$, so that
the maximally mixed state on the ground state subspace of
$H(\{s_X\}$ is $Z(\{s_X\})^{-1} \prod_{<a,b>} (I-Q_{a,b}) \prod_{a} (I-Q_a)$, where $Z(\{s_X\})$ is a normalization factor.
This state is
a $(2,0)$ trivial state on the coarse-grained lattice of regions because for any choice of the variables
$\alpha(a)$ for each region $a$ the projection of $\rho(\{s_X\})$ onto the product of spaces ${\cal H}_a^{\alpha(a)}$
is a product state on the spaces
${\cal H}_{a\rightarrow b}^{\alpha(a)} \otimes {\cal H}_{b \rightarrow a}^{\beta(b)}$.
Thus it is an $(R,0)$ trivial state on the
original lattice for $R=2 l_\beta$.  

We now bound the contribution of invalid configurations.
Note that given any two sequences $\{s_X\}$ and $\{s'_X\}$ such that $s'_X\geq s_X$ for all $X$ we have $Z(\{s'_X\}) \leq Z(\{s_X\}$.  Imagine breaking each large square of linear size $l_\beta$ into small squares of size $2R_{int}$ on each side (this will not give the best estimates but simplifies the proof),
 and let $n_C$ be the maximum number of projectors that intersect any of those small squares.
Then, the sum of
$Z(\{s_X\}) P(\{s_X\})$ 
over all $\{s_X\}$ such that there is a given configuration of holes in those small squares with a total of $k$ such holes
is at least equal to $\exp(-(2R_{int}^2 k\beta)$ times the sum of the same quantity over configurations with no holes.
So, the sum of $Z^{-1} Z(\{s_X\}) P(\{s_X\})$ over configurations with no holes in any given square
is bounded by
$( 1-\exp(-R_{int}^2 \beta))^{(l_\beta/R_{int})^2}$,
as follows by counting the number of configurations with holes and without.

Pick
\be
\label{lbeta}
l_{\beta}=\exp((2R_{int})^2 \beta) R_{int} \log(V)/\log(\epsilon),
\ee
so
$( 1-\exp(-(2R_{int})^2 \beta)^{(l_\beta/R_{int})^2} \leq \epsilon/V$.
Then, the contribution of configurations such that at least one large square has no holes to the trace in Eq.~(\ref{splitdefn})
is at most $\epsilon$.  Thus, for such an $l_\beta$ we can ignore such configurations and restrict to a sum over valid configurations giving an approximation to $\rho$ with error at most $\epsilon$ in trace norm, describing $\rho$ as an incoherent sum over trivial states; if desired, one
can write this sum in the form of Eq.~(\ref{trivdens}).

{\it Operational Properties---}
Consider first $T=0$.  The four dimensional toric code on a torus has a degenerate ground state. 
There
are surface operators $U_x$ and $U_z$ that act like the Pauli operators $\sigma_x$ and $\sigma_z$ on a two-dimensional subspace of the ground state space.  These operators anti-commute so that $\{U_x,U_z\}=0$.  Further, defining $U'_x$ and $U'_z$ to denote the
same operators translated a distance $L/2$ perpendicular to the given surfaces, the ground state expectation value of $U'_x U_x^\dagger$ equals  $1$, and similarly for $U'_z U_z^\dagger$.
At nonzero temperature,
the possibility of correcting errors\cite{4dtoric} implies that we can ``thicken"\cite{thicken} those surfaces, giving unitary operators $V_x$ and $V_z$ which are supported within some distance (say, $L/8$) of the given surface which have similar properties at $T>0$ to the operator $U_x$ and $U_z$ at $T=0$.  In particular, we expect that 
for sufficiently small $T$
\be
\label{incon1}
{\rm tr}(\rho V_x V_z V_x^\dagger V_z^\dagger)\approx -1,
\ee
\be
\label{incon2}
{\rm tr}(\rho V'_x V_x^\dagger)\approx 
{\rm tr}(\rho V'_z V_z^\dagger)\approx 
1.
\ee
However, Eqs.~(\ref{incon1},\ref{incon2}) are inconsistent
with having an $(R,\epsilon)$ trivial state for $R$ sufficiently small compared to
 $L$ and $\epsilon$ of order unity: 
if $\rho$ is an incoherent mixture of states $\psi_{triv}(a)$,
${\rm tr}(\rho V'_x V_x^\dagger)=\sum_a P(a)
\langle \psi_{triv}(a)|V'_x V_x^\dagger|\psi_{triv}(a)\rangle$.
Then, since the
separation between the supports of $V'_x$ and $V_x^\dagger$ is greater than
$R$, this equals
$\sum_a P(a) \langle \psi_{triv}(a)|V'_x|\psi_{triv}(a)\rangle \, \langle \psi_{triv}(a)| V_x^\dagger|\psi_{triv}(a)\rangle$.  For this sum to be close to $1$ as in Eq.~(\ref{incon2}), a $\psi_{triv}$ chosen at random must be, with probability close to $1$, an approximate eigenstate of both $V_x$ and $V_z$.  However, 
this is inconsistent with Eq.~(\ref{incon1}).
One can derive a similar inconsistency result for certain defect creation processes similar to those in \cite{localnotes}.

Conversely, we can comment on this usefulness of a trivial state for storing quantum information.  To manipulate quantum information
at $T>0$, we need operators $V_x$ and $V_y$ which act like the Pauli matrices on a single qubit, as in Eq.~(\ref{incon1}).
However, since Eqs.~(\ref{incon1},\ref{incon2}) are inconsistent with having a trivial state, we cannot have also corresponding operators $V'_x,V'_y$.  Further, such a trivial mixed state can be no more useful for storing information than a trivial pure state.
Note that it seems too much to hope for the converse statement (that a nontrivial state is useful for storing quantum information), since even at $T=0$ the Chern insulator\cite{chernI} provides an example of a circuit nontrivial state that has a unique ground state and so cannot be a quantum memory.

{\it Discussion---}
We have proposed a definition of topological order $T>0$.  While this definition is simple, it allows us to make precise statements about how quantum information can be manipulated in two dimensional Hamiltonians which are sums of commuting terms.
This raises the question of what can happen for arbitrary two dimensional Hamiltonians.
In three dimensions, we expect that discrete gauge theories are topologically trivial at $T>0$ under the circuit definition (see Appendix).
A much more interesting question is whether Haah's code\cite{haah1,haah2}, which avoids this gauge theory paradigm, is trivial or not at $T>0$.
Using quasi-adiabatic continuation\cite{qad}, we can relate the circuit definition at $T=0$ to whether or not one can deform one local Hamiltonian into another while avoiding a phase transition; however, an analogous question for $T>0$ is open (whether the absence of a  phase transition in thermodynamic quantities when lowering the temperature from infinity to some finite $T_f$  implies that the density matrix at $T_f$ is trivial).
Finally, we ask if
there is a circuit definition for exotic critical points\cite{deconcp}.

{\it Acknowledgments---} I thank D. Poulin for useful comments.

\newpage
\appendix
\section{An Example Hamiltonian With Topological Order Only at $T>0$}
Consider the following Hamiltonian.  It is defined on a four dimensional lattice.  On each plaquette of the lattice, we define a spin-$1/2$ degree of freedom.
On each site, we define a degree of freedom with $3$ different states, labelled $|0\rangle, |1\rangle, |2\rangle$.  Define an interaction term for the site degrees of freedom, $H_s$, as follows:
\begin{eqnarray}
H_s&=&-J \sum_{<i,j>} |0\rangle_i \langle 0| \otimes |0 \rangle_j \langle 0| \\ \nonumber
&&-J \sum_{<i,j>} \Bigl(|1\rangle_i \langle 1|+ \otimes |2 \rangle_i \langle 2| \Bigr)\otimes  \Bigl(|1\rangle_j \langle 1|+ \otimes |2 \rangle_j \langle 2| \Bigr) \\ \nonumber
&&+h\sum_{i} |0\rangle_i \langle 0|.
\end{eqnarray}
This Hamiltonian is classical, in that it is diagonal in the given basis.  We take $J>0$.  The sums are over nearest neighbor $i$,$j$.  The first term favors having $0$ on site $i$ neighboring $0$ on site $j$.  The second term favors having $1$ or $2$ on site $i$ neighboring $1$ or $2$ on site $j$.  The third term favors having $0$ on each site for $h<0$ and favors $1$ or $2$ for $h>0$.
The ground state has all sites in the $0$ state, but one can tune the interaction terms to produce a transition at non-zero temperature to a state where most spins are in the $1$ or $2$ state (since one can choose each spin to be in either the $1$ or $2$ state, there is an increase in entropy which can outweigh the $h$ term in the Hamiltonian), with just a few excitations to the $0$ state (an isolated $0$ is penalized in this state, since most of its neighbors are $1$ or $2$; it is only when many $0$ are nearby that one gains energy).  The transition is likely first order.

Now add interaction terms for the plaquette degrees of freedom.  The toric code in four dimensions has two terms, which we call $A$ and $B$, which act on plaquettes either surrounding a link or in a cube.  These terms are diagonal in either the $X$ or $Z$ basis respectively.  Define new interactions terms $C$, $D$ as follows: $C$ is equal to $A$ multiplied by the product, over sites on that link, of the projectors onto the space spanned by the $1$ and $2$ states on those sites and $D$ is equal to $B$ multiplied by the product, over sites in that cube, of the projectors onto the space spanned by the $1$ and $2$ states on those sites.  Finally, we add an additional term $E$ which is equal to the sum over plaquettes of $\sigma^z$ on that plaquette multiplied by the product, over sites bordering that plaquette, of the projector onto the the $0$ state on those sites.

Then, we define the Hamiltonian
\be
H_s+C+D+E.
\ee
By tuning $h$, we can achieve the following phase diagram as a function of temperature: the ground state has $0$ states on all sites.  Then, the $C,D$ terms are equal to zero, and the $E$ term polarizes the plaquette degrees of freedom in the $Z$ direction.  Hence, the ground state is a trivial product state.

However, as we increase the temperature, this changes.  Consider the state which is maximally mixed over $1$ and $2$ degrees of freedom on the sites, multiplied by the toric code ground state.  This state gains entropy (due to the mixture of $1$ or $2$ states).  This state has a different energy from the ground state.  Part of the difference arises from the differing $h$ term, and part of the difference is due to the different energy of the expectation value of $C+D+E$ in those two states.  However, by tuning $h$ we can make this state have very slightly higher energy density than the ground state.
Thus, it seems very likely that we can tune $h$ to produce a first order phase transition from the ground state at zero temperature to a state at some small $T>0$ which contains the four dimensional toric code.  Now, this state at non-zero temperature is not exactly the same as the toric code at non-zero temperature, since any $0$ states on the sites will change the interaction terms in the toric code Hamiltonian nearby leading to some correlation between defects in the toric code state; however since those $0$ states are rare at low temperature (by tuning $h$ we can accomplish the phase transition between states at a very low temperature to keep the $0$ states rare), it is likely that the effect of these $0$ states will not destroy the topological order in the toric code Hamiltonian.

\begin{figure}
\includegraphics[width=1.9in]{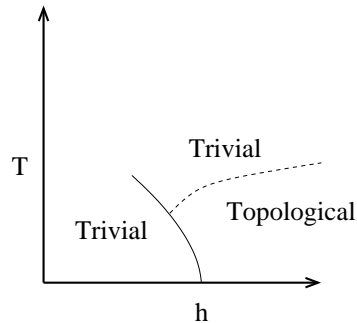}
\caption{Illustration of phase diagram expected for the toy model.  The topological phase is the topologically ordered phase of the four dimensional toric code.  The trivial phases are different trivial phases.  The solid line indicates a first order transition.  The dashed line indicates a phase transition in which the site degrees of freedom remain primarily in the $|1\rangle$ or $|2\rangle$ states, but the toric code portion of the system becomes topologically trivial.  Scales of axes are arbitrary, as is the particular value of $h$ at $T=0$ at which the phase transition occurs.  The point of the model system is that a point at $T=0$ with $h$ slightly smaller than the critical $h$ can change from trivial to topological as temperature is increased.
}
\end{figure}

 Thus, while this system is topologically ordered at such temperatures under the circuit definition, any reference to ground state excitations cannot capture this.  There is a range of temperatures where one can store quantum information in a robust matter, but it is not possible at zero temperature for this $h$.  Error correction is accomplished by using the same error correction procedure as in the four dimensional toric code.  Of course, this example is quite artificial, but other examples may exist close to zero temperature phase transitions: if one is close to a first order phase transition at zero temperature, as we are here with the phase transition being driven by changing $h$, then increasing temperature can lead to a situation where the situation at a nonzero temperature for a given value of $h$ (a value that favors the $0$ state at zero temperature) is in the same phase as the zero temperature state at a different value of $h$ (a value that favors the $1$ or $2$ states).
The figure illustrates the phase diagram for this model; more realistic models with similar phase diagrams probably exist.

\section{Triviality of Three Dimensional Toric Code at $T>0$}
We now show the triviality of the three dimensional toric code at $T>0$.  It is possible that this method could be extended to show the triviality of
other discrete gauge theories in three dimensions; our approach relies on constructing operators with certain commutation relations but probably similar operators exist for other discrete gauge theories.

The three dimensional toric code has topological order at $T=0$.  It has a phase transition at non-zero temperature, and the topological entropy
becomes non-vanishing below this phase transition temperature\cite{topomi}.

The Hamiltonian is defined by a three dimensional cubic lattice, with spin-$1/2$ degrees of freedom on the links of the lattice.
We define two operators, plaquette and star operators, by
\be
B_p=\prod_{i\in p}\sigma^z_i,
\ee
\be
A_s=\prod_{i\in s} \sigma^x_i.
\ee
These operators commute and the Hamiltonian is defined by
\be
H=-\lambda_A \sum_s A_s - \lambda_B \sum_p B_p.
\ee

Much of the three dimensional case follows the two dimensional case.
Similar to Fig.~1 in two dimensions, in three dimensions we divide the cubic lattice into cubes of linear size $l_\beta$.  We again write the thermal density matrix $\rho$ as a sum over $\rho(\{s_X\})$, as in Eq.~(\ref{sumup}).  As in two dimensions, for $l_\beta$ taken to be of a size which is of order $\log(V)/\log(\epsilon)$,  one can show that the sum is dominated by ``valid" configurations, where now a configuration is valid if there is at least one hole per cube.

However, at this point the three dimensional result needs new techniques.  In two dimensions, we showed that each $\rho(\{s_X\})$ for any valid configuration was trivial using techniques based on ideas from \cite{bv} which work for any Hamiltonian which is a sum of commuting projectors.  However, in three dimensions, we need to do something different, since the Hamiltonian $H(\{s_X\})$ {\it cannot} be coarse-grained to a Hamiltonian in which each term acts on only at most two sites (we would need to have long ``line-like" holes to be able to do this whie the holes are only small ``point-like" configurations here).  So, we develop a technique that works for the three dimensional toric code to show that $\rho(\{s_X\})$ is trivial for valid configurations.  This method relies on constructing particular operators using Pauli spin operators of the toric code.

Consider the operator that we call $F_{i,s}$, where $i$ labels a bond and $s$ labels a site that neighbors the bond $i$.
This operator is defined by
\be
F_{i,s}\equiv \sigma^z_i A_s.
\ee
That is, $F_{i,s}$ is the product of $i\sigma^y$ on site $i$ with $\sigma^x_j$ over sites $j \neq i$ with $j\in s$.
We consider, as mentioned, the case that $s$ neighbors $i$.  Thus, there are two possible choices of $s$ for any given $i$, and those choices can be thought of as defining a direction: $s$ is defined by the bond $i$ and the direction that one must move on the bond $i$ to arrive at site $s$.

Note that $F_{i,s}$ commutes with all operators $B_p$.  Also, $F_{i,s}$ commutes with all operators $A_t$ {\it except} that it anticommutes with the two operators $A_t$ with $t$ neighboring $i$ (one of these operators $A_t$ is equal to $A_s$ while the other is obtained by moving on the opposite direction on bond $i$).
Define the unitary
\be
U_{i,s}\equiv \exp(i \frac{\pi}{4} F_{i,s}).
\ee
Because of the commutation relations above, this unitary obeys:
\be
\label{UB}
U_{i,s} B_p U^\dagger_{i,s}=B_p,
\ee
and
\be
\label{UA}
i \not \in t \; \rightarrow \; U_{i,s} A_t U^\dagger_{i,s}=A_t,
\ee
where the notation $i \not \in t$ means that bond $i$ is not a neighbor of site $t$.
However,
\be
\label{UotherA}
U_{i,s} A_s U^\dagger_{i,s}=\sigma^z_i.
\ee

We consider a general class of Hamiltonians
\be
H({\cal S},{\cal P},{\cal B})=-\lambda_A \sum_{s\in {\cal S}} A_s - \lambda_b \sum_{p \in {\cal P}} B_p - \lambda_A \sum_{i \in {\cal B}} \sigma^z_i,
\ee
where ${\cal S},{\cal P},{\cal B}$ are sets of vertices, plaquettes, and bonds, respectively.  Thus, the three dimensional toric code Hamiltonian $H$ is $H({\cal S},{\cal P},{\cal B})$, with ${\cal S},{\cal P}$ being the sets of all vertices and plaquettes respectively, while ${\cal B}$ is the empty set.

Every configuration of the $\{s_X\}$ defines a Hamiltonian $H({\cal S},{\cal P},{\cal B})$, with ${\cal B}$ being the empty set, and ${\cal S},{\cal P}$ being some subset of the vertices and plaquettes, depending upon which $s_X$ are non-zero.
Now, consider a configuration of the $\{s_X\}$ and the corresponding Hamiltonian $H({\cal S},{\cal P},{\cal B})$.  

We define a certain unitary transformation for each closed surface on the dual lattice.
A surface on the dual lattice is made of plaquettes on the dual lattice and each such plaquette intersects some spin-$1/2$ degree of freedom.
We assume our surface is orientable (this will hold for all the surfaces described below, which are contractible, but will exclude perhaps certain exotic situations on certain three dimensional manifolds), and we imagine an arrow on each bond which intersects the surface; this arrow will point in the direction of the orientation.
For each such surface, and each bond $i$, let $s(i,\xi)$ be the vertex reached by starting on bond $i$ and moving in the direction of that arrow.
We will be interested in operators $U_{i,s(i,\xi)}$ below.  Let $s(i,\overline \xi)$ be the vertex reached by starting on bond $i$ and moving {\it opposite} to the direction of the arrow.

Let us say that this surface is a ``free surface" if for all bonds $i$ which intersect the surface, the operator $A_{s(i,\overline \xi)} $ does not appear in the Hamiltonian $H({\cal S},{\cal P},{\cal B})$ and if none of the operators $F_{i,s(i,\xi)}$ intersect any bond in ${\cal B}$.
Note that if $s_X$ contains a hole, then the surface surrounding that hole, oriented outwards, is a free surface for $H({\cal S},{\cal P},{\cal B})$.

Take any valid Hamiltonian $H(\{s_X\})$, take any free surface $\xi$  surrounding a hole and orient that surface outwards.   Take any operator $U_{i,s(i,\xi))}$ for any $i$ intersected by the free surface.  
Consider the Hamiltonian $U_{i,s(i,\xi)} H(\{s_X\}) U_{i,s(i,\xi)}^\dagger$.  This define some new Hamiltonian; the plaquette terms $B_p$ are unchanged by this producedure, as all are all the vertex terms $A_s$ except for the vertex term $A_s$ for $s=s(i,\xi)$ which is turned into the operator $\sigma^z_i$ by Eq.~(\ref{UotherA}).  Note that since the surface is free, there are no $A_s$ intersecting the surface on the inside of the hole (such an operator occurs for $s=s(i,\overline \xi)$ which does not appear by definition of a free surface); the only such $A_s$ are those on the outside of the hole.
Note that this new Hamiltonian $U_{i,s(i,\xi)} H(\{s_X\}) U_{i,s(i,\xi)}^\dagger$ has a new free surface: this free surface is given by taking $\xi$ and moving it outwards along bond $i$.
Repeating this procedure, we can move each free surface outwards.  Note that since the original Hamiltonian has ${\cal B}=\emptyset$, all operators $\sigma^z_i$ that are produced by this procedure are produced far enough inside the free surfaces that they do not stop the surfaces considered in future steps of this procedure from being free; that is, such operators are never transformed by the operators $U_{i,s(i,\xi)}$ in later steps.

We then choose a specific order to apply this procedure: in the first round, we take each free surface surrounding each hole, and iterate this procedure until that free surfaces reach the boundary between cubes.  This first round (which is the first round of the quantum circuit that we use to show that $\rho(\{s_X\})$ is trivial) is a product of unitaries, one on each cube, each such unitary being the product of the $U_{i,s(i,\xi)}$ used to move that surface out until it hits the boundary.  Then, in the next round, we move the free surfaces one further step outwards, so that they intersect; once the free surfaces intersect, the terms $A_s$ are removed from the Hamiltonian.
The final Hamiltonian contains only terms diagonal in the $S^z$ basis, and hence is classical.  Thus, the original Hamiltonian $H(\{s_X\})$ is trivial.

\begin{thebibliography}{99}
\bibitem{thicken} Given an algorithm that measures the error syndrome and corrects errors using local unitaries,
construct
a unitary $W$ that does the same error correction while storing the state of the syndrome in some previously initialized auxiliary qubits; $W^\dagger U_x W$
acts as a unitary on the original system if the auxiliary qubits are properly initialized.  Let $V_x$ be the closest approximation to this unitary among all unitaries supported on sites within distance $L/8$ of the support of $U_x$.  Eqs.~(\ref{incon1},\ref{incon2}) follow if the approximation error is small; this error is indeed expected to be small for an appropriate algorithm in the topologically ordered phase as with probability close to $1$ the error correction near the support of $V_x$ is independent of the syndrome far away.
\end{thebibliography}
\end{document}